\documentclass[usenatbib]{aa}

\usepackage{graphicx}
\usepackage{amssymb}
\usepackage{epsfig}
\usepackage{psfig}
\usepackage{natbib}
\bibpunct{(}{)}{;}{a}{}{,}

\def\pl{{\sc pl}}

\def\bb{{\sc bb}}

\def\bb{{\sc bb}}

\def\gx{GX~3+1}
\def\be{\begin{equation}}
\def\ee{\end{equation}}

\begin{document}
\title{Two-phase X-ray burst from GX 3+1 observed by {\em INTEGRAL}
}

\author{J. Chenevez \inst{1} \and
        M. Falanga \inst{2,3} \and
        S. Brandt \inst{1} \and
        R. Farinelli \inst{4} \and\\
        F. Frontera \inst{4,5} \and
        A. Goldwurm \inst{2,3} \and
	J.J.M. in 't Zand \inst{6,7} \and
        E. Kuulkers \inst{8} \and
        N. Lund \inst{1}}

\offprints{J. Chenevez \\\email{jerome@spacecenter.dk}}
\titlerunning{Two-phase X-ray burst from GX 3+1}
\authorrunning{J. Chenevez, M. Falanga  et al.}

\institute{Danish National Space Center, Juliane Maries Vej 30, DK-2100
           Copenhagen Ø, Denmark
  \and     CEA Saclay, DSM/DAPNIA/Service d'Astrophysique (CNRS FRE
           2591), F-91191, Gif sur Yvette, France
  \and     Unit\'e mixte de Recherche Astroparticule et
           Cosmologie, 11 Place Berthelot, F-75005 Paris, France
  \and     Dipartimento di Fisica, Universit\`a di Ferrara, via Saragat 1, I-44100 Ferrara, Italy
  \and     Istituto di Astrofisica Spaziale e Fisica Cosmica
           CNR, via Gobetti 101, I-40129 Bologna, Italy
  \and     SRON Netherlands Institute for Space Research, Sorbonnelaan 2, 3584 CA Utrecht, The Netherlands
  \and     Astronomical Institute, Utrecht University, PO Box 80 000, 3508 TA Utrecht, The Netherlands
  \and     ESA/ESAC, Urb. Villafranca del Castillo,
           P.O. Box 50727, E-28080 Madrid, Spain
}

\abstract{
{\em INTEGRAL} detected on August 31, 2004, an unusual thermonuclear X-ray
burst from the low-mass X-ray binary GX~3+1. Its duration was 30~min,
which is between the normal burst durations for this source
($\la10$~s) and the superburst observed in 1998 (several hours).
We see emission up to 30 keV energy during the first few seconds of the
burst where the bolometric peak luminosity approaches the Eddington limit. 
This peculiar burst is characterized by two distinct phases:
an initial short spike of $\sim$6~s consistent with being similar to a normal type I X-ray burst, 
followed by a remarkable extended decay of cooling emission. 
We discuss three alternative schemes to explain its twofold nature: 
1) unstable burning of a hydrogen/helium layer involving an unusually large amount of hydrogen,
2) pure helium ignition at an unusually large depth (unlikely in the present case), and
3) limited carbon burning at an unusually shallow depth triggered by unstable helium ignition.
Though none of these provide a satisfactory description of this uncommon event, the former one 
seems the most probable.



}
\maketitle


\section{Introduction}
\label{sec:intro}

Many of the observed low-mass X-ray binary systems are known to exhibit
type I X-ray bursts. The X-ray light curves of such events are
characterized by a fast rise time followed by an exponential decay.
The decay time varies from burst to burst, but is generally between a
few seconds and a few minutes. For a given burst the decay times are
shorter at higher energies. They are produced by unstable burning of
accreted matter on the surface of the neutron star. The emission can
be described well by black-body radiation with temperatures, $kT$, in the
range of a few keV. The energy dependent decay time of these bursts
is attributed to the cooling of the neutron star photosphere resulting
in a gradual softening of the burst spectrum.
For a review, see, e.g., \citet{lewin93, sb03}.

\gx\ is a well-known X-ray burster, exhibiting exclusively short
($\la10$~s long) bursts \citep[see][and references therein]{hartog03} 
except once when it exhibited an hours-long so-called
'superburst' \citep{k02}. Superbursts are thought to arise from
carbon shell flashes 
in the layers below the surface\citep[e.g.,][]{cb01}. One short burst
was shown to exhibit photospheric radius expansion from which a
distance of $\sim$5~kpc could be estimated \citep{kvdk00}.

In this letter we present an analysis of a second unusual burst from
\gx\ which was observed on August 31, 2004 with the {\it International
Gamma-Ray Astrophysics Laboratory} ({\em  INTEGRAL}) and first
reported by \citet{blc04}.

\begin{figure}[h]
\begin{center}
\epsfig{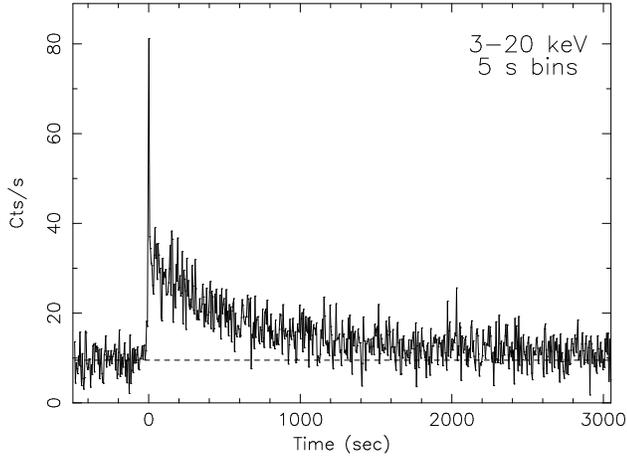}
\caption[]{Long X--ray burst from \gx\ on August 31, 2004.
The time zero corresponds to UTC 18:55:11.}
\end{center}
\label{fig:global_jmx}
\end{figure}


\section{Observations and Data Analysis}
\label{sec:integral}

The data were obtained with {\it INTEGRAL} \citep{w03}
between August 30 and September 3, 2004.
The observation was divided into a sequence of pointings separated by
${\sim}2^\circ$ on the sky. Each pointing had a duration of one hour.
For the X-ray monitor, {\em JEM-X}, this pointing strategy implies that
the visibility of any given source changes significantly every hour.
At the time of the burst the {\em JEM-X} effective area for \gx\ was
about 60\% of the on-axis value. In the previous pointing it had been
only 15\% and in the following one it increased to 95\%.
In our analysis we have used data from the ``burst" pointing and
the following pointing (``science windows" 9 and 10 from
{\it INTEGRAL} revolution 230).

We use data from {\em JEM-X} \citep{lund03} from 3 to 25 keV,
and from {\em IBIS/ISGRI} \citep{u03,lebr03} at energies between 18
and $\sim$50 keV. Data reduction and analysis were performed using
the standard Offline Science Analysis ({\em OSA}) v.5.0 \citep{c03}
and public {\em JEM-X} specific software \citep{lund04}.

All the light curves in this paper are based on events selected
according to the detector illumination pattern for \gx .
For {\em ISGRI} we selected ``source"-events using an illumination
threshold of 0.6; for {\em JEM-X}, which utilizes a different type
of mask \citep{lund03}, we used a lower threshold of 0.25.


\section{Results}
\label{sec:res}

\subsection{Burst light curves}

Figure 1 displays
the {\em JEM-X} light curve in the 3 to 20 keV
band with a time resolution of 5~s for the full science window
in which the unusual burst from \gx\ was observed.
The burst consists of two distinct phases:
an initial short spike and an extended decay phase.
Figure 2 shows the initial spike with 0.5 s resolution for a
number of energy bands. The spike emission is well visible
all the way up to 30 keV. The 18-30 keV band from {\em ISGRI}
light curve shows indications of a double peak structure
during the spike phase.
We have verified that the background count rate is stable over
the full science window in which the burst is observed. We can
therefore exclude interference from other time variable sources
within the field-of-view.


\begin{figure}
\begin{center}
\epsfig{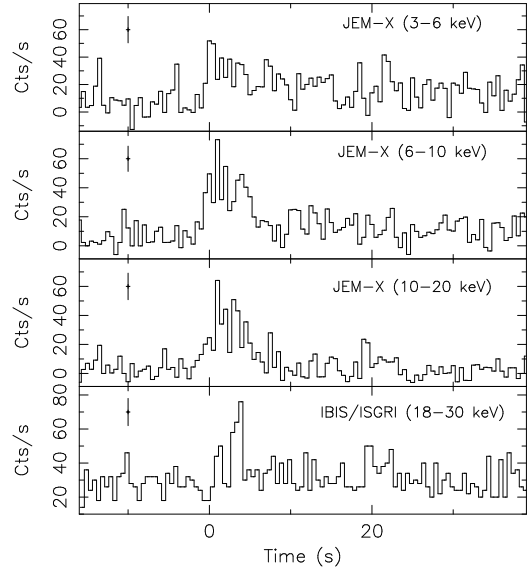}
\caption[]{Light curves of the initial part of the X-ray burst
with 0.5 sec binning in selected energy bands.
A reference error bar is shown in the corner of each panel.
The background has been subtracted from the {\em JEM-X}
light curves, but not from the {\em ISGRI} one.}
\end{center}
\label{fig:lc_isgri}
\end{figure}



\subsection{Spectral Analysis}
\label{sec:spectrum}

We have performed a time resolved spectral analysis based on the
{\em JEM-X} data ({\em ISGRI} detected \gx\ only for a few seconds
during the spike phase of the burst and for such a short time the
data are inadequate for a reliable spectral analysis).
The energy range covered by {\em JEM-X} does not allow us to constrain
well the interstellar column density N$_{\rm H}$. Therefore, we have
fixed in all our spectral fits the N$_{\rm H}$--value at 
1.6$\times 10^{22}$ atoms ~cm$^{-2}$ as derived by \citet{Oosterbroek01}.

\begin{figure}{}
\begin{center}
\epsfig{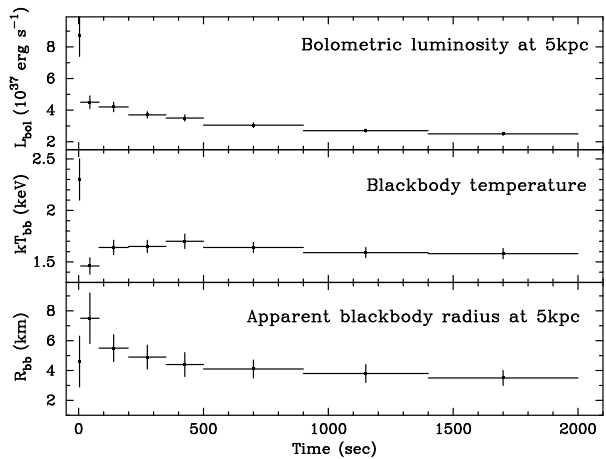}
\caption{Results of the time resolved spectral analysis.}
\label{fig:spec_evolution}
\end{center}
\end{figure}

We have determined the spectral evolution during the burst and
compared the burst spectra to the persistent source spectrum
which we extracted from the subsequent science window.
The burst is divided into time intervals as shown in
Fig. \ref{fig:spec_evolution}, the shortest interval being
$\sim$8~s during the spike.
We defined the energy channels in the burst spectral analysis such
to assure a uniform distribution of counts in all spectral channels.
Following an approach proposed by \citet{vPL86}
\citep[see also][]{Sztajno}, we have modelled the total burst emission
averaged during each time interval by using the same two spectral
components as for the persistent emission. The model contains a
black-body (\bb) component for the thermal emission and a power-law (\pl)
component for the Comptonized photons. 
The power-law component is thought to originate
far from the neutron star - in the accretion disk environment - and
consequently is not expected to be strongly affected by the events on
the neutron star. Therefore we determine the power-law contribution
through a fit to the persistent emission, and then keep this component
fixed  throughout the analysis of the burst spectra.
The upper panel of Fig. \ref{fig:spec_evolution} displays the bolometric
luminosity (0.1-200 keV) of the source obtained from the two--component
spectral model assuming a distance of 5~kpc.
The inferred black-body temperature and apparent
black-body radius are shown in the middle and lower panels,
respectively. The softening of the emission towards the end of
the decay phase is also indicated by the e-folding 
decay times going from 1110 $\pm$170~s in 
3--6 keV to 510 $\pm$160~s in 10--20 keV.
The best fit parameters for the persistent emission and for the spike
are given in Table \ref{tab:spec} as well as usual burst parameters.


We note that a two-component spectral analysis of the total burst
emission may not always be an accurate approach \citep{Ketal02}.
Therefore, we also checked that the net burst emission (by subtracting
the persistent emission during the burst) is satisfactorily described
by a black-body model, and repeating the spectral analysis we found the
same trend showing a maximum luminosity only a factor 1.3 lower, and a
slightly deeper decrease of the temperature (by a factor 1.2)
simultaneously with a slightly higher increase of the radius
(by a factor 1.3).


\begin{table}
\begin{minipage}[t]{\columnwidth}
\caption{Burst analysis results}
\label{tab:spec}
\centering
\renewcommand{\footnoterule}{}
\begin{tabular}{lcc}
\hline\hline
Dataset ({\em JEM-X}) & Persist. & Spike \\
Model           & \bb\ \& \pl\ & \bb\ \& \pl\ \\ \hline \vspace{1mm}
N$_{\rm H}$\footnote{Parameter fixed; see text.} ($10^{22}$ atoms cm$^{-2}$) & 1.6  & 1.6\\
$kT_{\rm bb}$ (keV)  & 1.7$^{+0.1}_{-0.1}$ & 2.3$^{+0.5}_{-0.4}$\\
\vspace{1mm}
$R_{\rm bb}$ (km)    & 2.6$^{+0.6}_{-0.3}$ & 4.6$^{+1.9}_{-1.5}$\\
$\Gamma$   & 3.6$^{+0.5}_{-0.5}$ &  3.6 (fixed)\\
$L_{\rm (5-25 keV)}$ (10$^{37}$ erg s$^{-1}$) & 0.73 & 5.9 \\
$\chi^{2}/{\rm dof}$   & 88/76 &  27/29 \\
\hline
Burst parameters \\
$L_{\rm peak}$\footnote{2s spike maximum luminosity (erg s$^{-1}$); see text.} & $E_{\rm b}$\footnote{Net burst fluence (erg).}  & $\gamma$\footnote{$\gamma \equiv L_{\rm pers}/L_{\rm peak}$.} \\
$1.6 \times 10^{38}$ & $2.1 \times 10^{40}$ & $\simeq 0.14$ \\
\hline
\end{tabular}
\end{minipage}
\end{table}


\subsection{{\em ASM} observations}

\begin{figure}[hb]
\epsfig{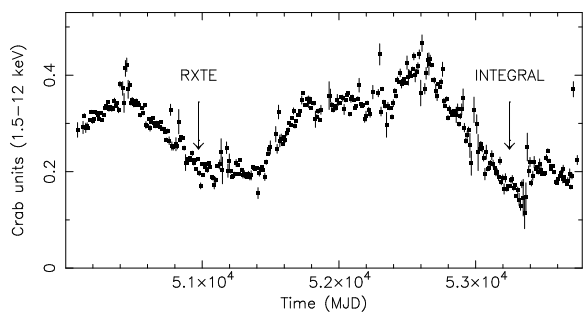}
\caption{{\em RXTE}/{\em ASM} light curve for \gx .
Arrows indicate {\em RXTE} superburst and {\em INTEGRAL} long burst.}
\label{fig:lc_asm}
\end{figure}

In Fig. \ref{fig:lc_asm} we show the 1.5--12 keV light curve from
August 1996 to December 2005 of GX~3+1. This light curve is based
on 10-day averages of the ``dwells'' executed by the {\em RXTE}/{\em ASM}.
The {\em ASM} count rate has been converted into flux using 1 Crab Unit
for 75~cts/s \citep [e.g][]{l96}. The persistent flux shows variations
of about a factor two on a $\sim$6~yr time scale.
Normal type I X-ray bursts have been observed at all intensity levels of
\gx\
\citep[][]{hartog03}.
We note that both our burst and the superburst observed by \citet{k02}
have been observed slightly before the persistent flux from the source
reaches its minimum. The persistent emission just prior to the
superburst (MJD 50973) was around 0.22~Crab, and 0.18 Crab prior to our
burst which occurred on MJD 53248.
Our burst was not seen by the {\em ASM}, so we cannot compare directly
the intensities of the two bursts in the {\em ASM} energy band.


\section{Discussion}
 \label{sec:conclusion}

Two distinct phases are evident in the burst from \gx\
observed on August 31, 2004: an initial hard spike of
duration $\sim$6~s followed by an extended quasi--exponential
decay phase lasting for more than 2000~s.

The initial spike has a rise time of 1.3 $\pm$0.1~s and fades away
with a decay time 3 $\pm$1~s.
The spike is similar in shape to normal type I burst from \gx\
of which almost hundred have been observed to date \citep{hartog03}.
The 18-30~keV emission of our burst shows a double peak
structure which may appear statistically marginal, but the main
part of this emission is clearly delayed with respect to the burst
onset. These characteristics could indicate a radius
expansion episode during the peak phase. We also note that the net-burst 
peak luminosity (see Table \ref{tab:spec}), as derived from a 2~s 
rebinned lightcurve during the spike phase, appears to be similar to the
peak luminosity observed for the radius expansion burst from \gx\
described by \cite{kvdk00}.

Only one previous observation exists for a radius expansion from GX~3+1, 
but the most unusual characteristic of our burst is the extended
decay phase lasting for more than 30 minutes. 
This is almost three orders of magnitude
longer than the normal type I bursts from GX~3+1. The rise
of the long lasting emission cannot be disentangled from the
spike emission, and we cannot say for sure which of the two
events triggered the other one.
We note that a similar kind of burst may have been observed from 
\object{Aql~X-1} \citep{Czerny87}. At the end of this observation $\sim$2500~s 
after the burst maximum the count rate was still $\sim$25\% higher than before the burst.
Other bursts with decay times of 20 to 30 minutes have occasionally been
observed from other sources
\citep[see e.g.][]{intZ02, Ketal02, mrls05}.
Some of these bursts also exhibited initial spikes, but these
spikes when present had softer spectra than the main burst
\citep [e.g.][]{Ketal02}. In our case the spike is much harder than
the emission during the decay phase. Our burst is also clearly 
different from the 1998 superburst in terms of duration,
decay time and total fluence.
%

Given that the total luminosity of the source during the first few
hundred seconds of the decay phase is less than half of the peak
luminosity we consider it unlikely that the source is still radiating
at the (hydrogen) Eddington luminosity during the decay phase
(this would require the source to be at 8 kpc). Moreover, it is
difficult in such a picture to reconcile the observed increase in
the emission temperature with the decrease in the bolometric flux (see Fig. 3).
Ideally an isotropically
emitting sphere supported by radiation pressure should maintain a
constant luminosity during the contraction phase. The observed
evolution may possibly be understood if the emission is not isotropic
but dominated by a 
band or a localized spot on the stellar surface. We have
searched for burst oscillations in the frequency range from 1 to 400~Hz
to confirm the notion of non-isotropic emission from a rotating neutron
star but have not detected any significant variations; we can put an upper limit 
of 5\% on the amplitude of a simple sinusoidal variation.

Assuming a source distance of 5~kpc we have estimated the persistent
bolometric (0.1--200 keV) luminosity of \gx\ to be
$L_{\rm pers}\sim2.2 \times 10^{37}$ erg s$^{-1}$.
We consider it safe to estimate the bolometric luminosity based on the
{\em JEM-X} data since we expect $\sim70\%$ of the flux to be contained within 4--25~keV range.
The corresponding persistent mass accretion rate is
$\dot M = L_{\rm pers}\;\eta^{-1}$
c$^{-2} \approx 1.2 \times 10^{17}$ g s$^{-1}$ (12\% $\dot M_{\rm Edd}$),
 where $\eta \sim$0.2 is the accretion efficiency for a neutron star.
This corresponds to an accretion rate per unit area
$\dot m = \dot M/A_{\rm acc}$ $\sim10^{4}$ g cm$^{-2}$ s$^{-1}$,
assuming a 10 km radius for the neutron star.

\gx\ has now been observed to produce three kinds of
type I X-ray bursts: normal, short helium flashes \citep[e.g.][]{hartog03},
one superburst \citep{k02}, and the present unusually long burst.
Bursts, with a duration more than 1000 s are rare, and
present theory provides three possible scenarios for such bursts -- the
burning of a large pile of hydrogen and helium, the burning of an 
unusually thick layer of helium \citep[see, e.g.,][]{intZ05, cum05}, 
or the carbon flash \citep[see, e.g.,][and references therein]{sb03}. 
%
%
The longevity of a mixed H/He flash is due to the hydrogen. While helium
fusion through triple-$\alpha$ burns the fuel within less than 1~s, the
rate of hydrogen fusion through rapid proton capture is limited by
slow $\beta$ decays \citep[see, e.g.,][and references therein]{sb03,lewin93}. 
Mixed H/He burning is expected to occur at
mass accretion rates below approximately 1\% and above 5\% of the
Eddington limit for a solar composition of the fuel. 
Between the two thresholds pure helium flashes occur which
result in $\sim$10~s bursts like usually observed in GX 3+1. The
lower threshold is related to the turn-off of stable hydrogen
burning. Thus, suddenly hydrogen becomes available for burning in
flashes and these are expected to become longer; also, the 
temperature drops which influences ignition conditions and burst
rates. 
Our burst appeared when the source went into a 10-yr intensity low 
at a luminosity close to $2 \times 10^{37}$ erg/s, which happens to coincide with 
a transition luminosity observed in other bursters \citep {cor03}. 
%
%
Usually, the burst durations increase
to at most a few hundred seconds, but sometimes they are similar as in
the present burst \citep{Czerny87}. The energetics of the burst are
consistent with this scenario. The radiated energy can be fueled by
solar composition material after accreting at 10\% of Eddington for
$\sim$9.4~h. No other bursts were detected in the 5.5~h continuous 
{\em INTEGRAL} observations preceding our burst.
In the nine days prior to the burst {\em JEM-X} accumulated $\sim$60~h exposure on GX~3+1
without detection of any burst which at least suggests that the burst
rate was low and accreted piles could become high enough.

The pure helium burning scenario is unlikely, because the appreciable 
accretion rate (0.12 times Eddington) results in such strong heating 
that the layer would flash before a sufficiently large thickness 
is reached to power such a long burst \citep{cum05}. 
Furthermore, the two-phased character of the burst
profile appears at odds with this scenario.

The third option, premature ignition of a carbon layer, may require
a special helium detonation geometry capable of producing a local
shock wave in the carbon layer below. The difficulty is that 
in order to achieve the observed decay time the carbon burning 
needs to occur much deeper than the helium layer. 
Compared to a superburst the duration is 
only consistent with a relatively small amount of carbon 
(few times $10^{10}$~gr~cm$^{-2}$ for a mass fraction $\la$0.1)
and the temperature 
needed to ignite the carbon might be too large \citep[see][]{cb01}. 

In conclusion, though we do not have a full and consistent picture of the 
conditions that led to the peculiar X-ray burst from GX~3+1,
we consider the mixed H/He scenario as the most likely one.


\begin{acknowledgements}
JC acknowledges financial support from the Instrument Center for Danish
Astrophysics.
MF acknowledges the French Space Agency and CNRS for financial support.
FF and RF acknowledge financial support from the Italian space agency ASI 
and Ministry of Education, University and Research (COFIN 2004).
\end{acknowledgements}

\bibliographystyle{aa}

\end{document}